\documentclass[a4paper]{spie}  
\usepackage[]{graphicx}
\usepackage{upgreek}   

\title{Gain-switched laser diode seeded Yb-doped fiber amplifier delivering 11-ps pulses at repetition rates up to 40-MHz}

\author{Manuel Ryser\supit{a}, Martin Neff\supit{a}, Soenke Pilz\supit{a}, Andreas Burn\supit{b} and Valerio Romano\supit{a,b}
\skiplinehalf
\supit{a}Institute of Applied Physics, University of Bern, Sidlerstr. 5, CH-3012 Bern, Switzerland; \\
\supit{b}Bern University of Applied Sciences, ALPS, Pestalozzistrasse 20, CH-3400 Burgdorf, Switzerland}

\authorinfo{Correspondence should be addressed to M.R. (email: manuel.ryser@iap.unibe.ch).}

\begin{document} 
\maketitle 

\begin{abstract}
Here, we demonstrate all-fiber direct amplification of 11 picosecond pulses from a gain-switched laser diode at 1063~nm. The diode was driven at a repetition rate of 40~MHz and delivered 13~$\upmu$W of fiber-coupled average output power. For the low output pulse energy of 0.33~pJ we have designed a multi-stage core pumped preamplifier based on single clad Yb-doped fibers in order to keep the contribution of undesired amplified spontaneous emission as low as possible and to minimize temporal and spectral broadening. After the preamplifier we reduced the 40~MHz repetition rate to 1~MHz using a fiber coupled pulse-picker. The final amplification was done with a cladding pumped Yb-doped large mode area fiber and a subsequent Yb-doped rod-type fiber. With our setup we achieved amplification of 72~dBs to an output pulse energy of 5.7~$\upmu$J, pulse duration of 11~ps and peak power of $>$0.6~MW.

\end{abstract}



\section{INTRODUCTION}
\label{sec:intro}  
In laser materials micro-processing ultra short laser pulses in the picosecond or femtosecond regime are used, when high demands on machining quality are posed. When processing metals, very good results concerning machining precision have been obtained with pulselengths in the range of 10~ps \cite{Jaeggi2011,Neuenschwander}. Under visual inspection, the lateral precision of the microprocessed details does not improve significantly, when reducing the pulselength down to 1~ps. However, recent studies show, that the efficiency of material removal significantly increases when processing is done with pulses in the sub-picosecond range \cite{Nedialkov2004}.

On the other side, when increasing the pulse length to above 10~ps, a slight precision decrease due to an increase in heat affected zone is noticed \cite{Hirayama2005,Zhu2000}.

This behavior is consistent with the observation that the ablation threshold fluence for metals decreases with decreasing pulselengths down to 10~ps and then does not noticeably change when further reducing the pulselength and some of its aspects are nicely explained by the two-temperature model if assuming an electron-phonon relaxation time of 10~ps \cite{Hirayama2005,Schmid2011}.

If ablation efficiency is to be considered, workers in the field show that material ablation with pulse durations of 10~ps and less are more efficient in terms of volume ablation rate compared to pulse durations of $>$20~ps \cite{Jaeggi2011}. For shorter pulse durations $<$10~ps, namely 0.1~ps-5~ps, it has been shown that Ablation of iron (Fe) becomes even more efficient.\cite{Nedialkov2004}. Roughly speaking, processing of metals with nanosecond pulse durations allow to control the spatial ablation resolution at a length scale of tens of micrometers. With pico- and femtosecond pulse durations material processing of structures on the nanometer length-scale can be achieved. This applies for the ablation resolution per pulse in depth. However, the achievable spot size of the beam is diffraction limited and thus in the order of the laser wavelength.

Considering these boundary conditions, an ideal laser system with large material removal rate while maintaining high ablation resolution must deliver short pulse durations (10~ps and shorter), high peak power ($\approx$MW) and thus relatively low pulse energy, good beam quality ($M^2\approx1$) and high pulse repetition frequency (MHz).

For these parameters we have conceived a compact fiber based system. In this work we present an all-fiber amplified gain-switched laser diode. It consists of a gain-switched laser diode and five amplification stages. The pulses from the gain-switched laser diode are temporally compressed in the first amplification stage by introducing appropriate dispersion compensation. Each amplification stage was optimized towards small contributions of amplified spontaneous emission (ASE) and small temporal pulse distortions. In order to reach high peak power without significant pulse shape distortions, the final amplification is done using an large mode area fiber (LMA) with a core radius of 30~$\upmu$m and a rod-type fiber with a core diameter of 70~$\upmu$m. 

To the best of our knowledge, we achieved the shortest pulse duration and highest peak power with a MHz repetition rate driven gain-switched diode based fiber amplifier system.

\section{Experimental Setup} 
\begin{figure}
   \begin{center}
   \begin{tabular}{c}
   \includegraphics[width=16.5cm]{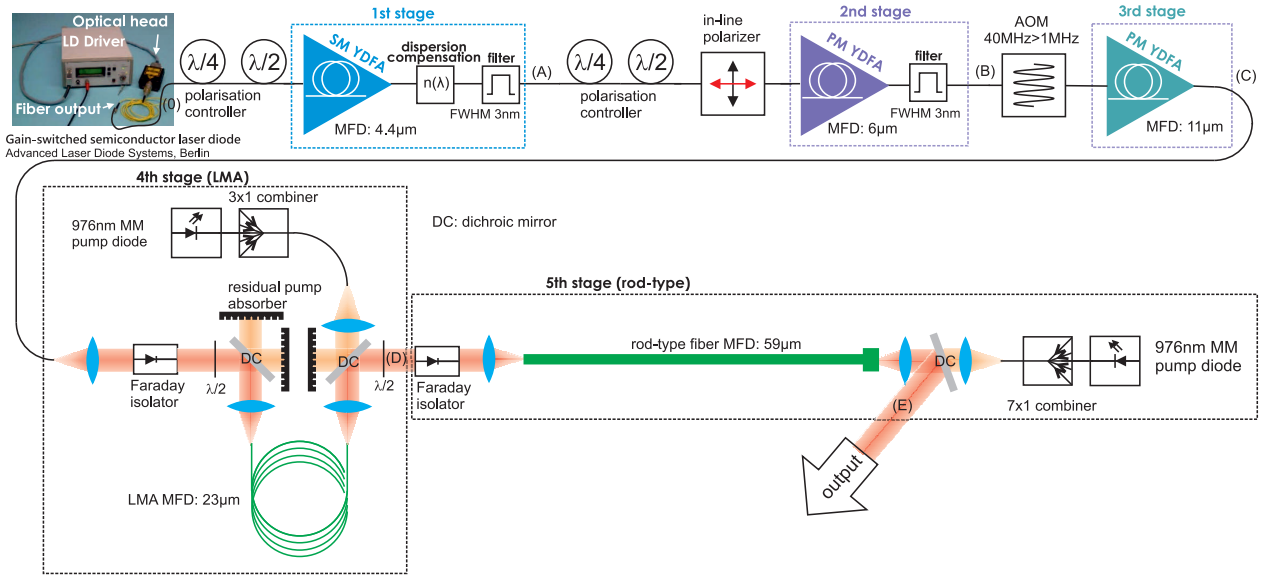}
   \end{tabular}
   \end{center}
   \caption[example] 
   { \label{fig:experimental_setup} Experimental setup. The gain-switched laser diode seeded fiber amplifier consisted out of five amplification stages.}
   \end{figure}

The gain-switched laser diode seeded Ytterbium-fiber amplifier was composed of five stages as sketched in Figure~\ref{fig:experimental_setup}.

\begin{figure}
   \begin{center}
   \begin{tabular}{c c}
   \includegraphics[width=7cm]{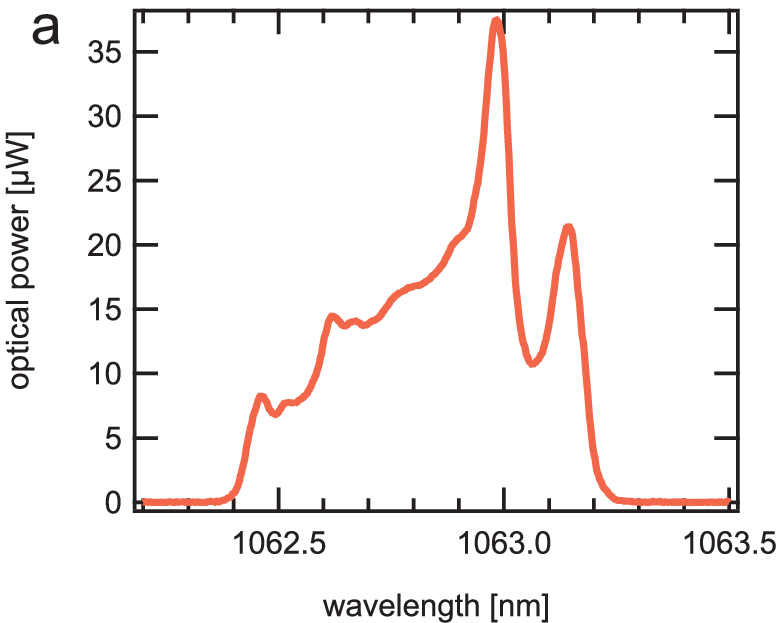} & \includegraphics[]{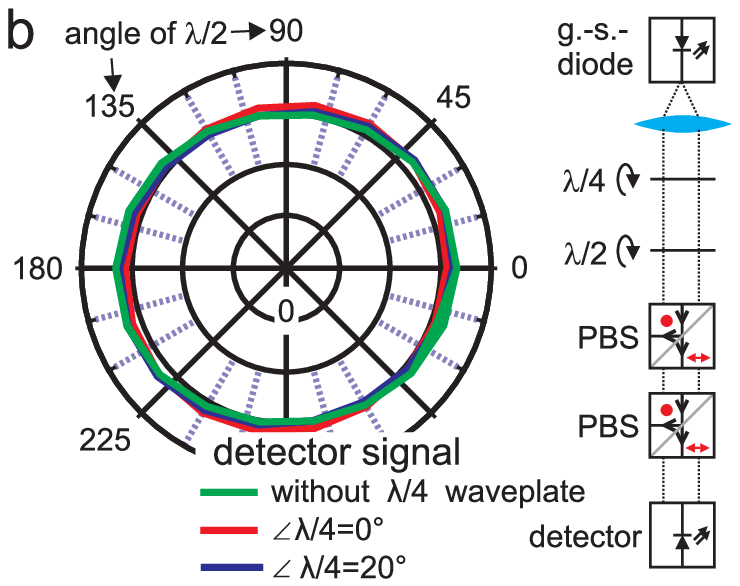} \\
   \end{tabular}
   \end{center}
   \caption[example] 
   { \label{fig:seed} Characteristics of the gain-switched laser diode. a) The maximum spectral emission of the gain-switched laser diode was at 1063~nm and the full bandwidth was 0.7~nm. b) Polarisation measurements of the gain-switched laser diode output (not fiber-coupled) showed that the output beam was unpolarized. }
   \end{figure}

As seed we used an unpolarized gain-switched laser diode (Advanced Laser Diode Systems, Berlin) with a maximum spectral output power at 1063~nm, a bandwidth of 0.7~nm (full width) and a fiber coupled seed power of 13~$\upmu$W at a fixed repetition rate of 40.03$\pm0.15$~MHz that is set by the driving electronics (c.p. Figure~\ref{fig:seed}). 

For the amplification we used commercially available Ytterbium-doped fibers. To reduce the nonlinear effects in the amplification stages, the fiber  mode-field diameters were increased for the successive stages.

The first pre-amplification stage consisted out of an unpolarized low noise ytterbium doped fiber amplifier (YDFA) in order to pre-amplify the weak unpolarized seed. Before and after the first stage we inserted polarization controllers to adjust the polarization plane that passes the in-line polarizer placed directly after the first YDFA-stage. For all amplification stages after the in-line polarizer were used polarization maintaining fibers.

In between the second and third YDFA-stage we inserted a polarization maintaining fiber-coupled acousto-optic modulator that was set to reduce the repetition rate from 40~MHz down to 1~MHz with a duty cycle of~$<$2.5\%. 

The first two YDFA stages were core-pumped at a wavelength 974~nm with Bragg grating stabilized single mode pump laser modules. 
The third, fourth (LMA) and fifth (rod-type) YDFA-stage were cladding pumped with a multi-mode pump module. Each multi-mode pump diodes was wavelength stabilized at 976~nm by controlling the temperature with a Peltier element and a TEC-controller. Each diode delivered a maximum optical output power of 8W. For the fourth YDFA-stage (LMA) we combined three pumps with a 7x1 multi-mode combiner and for the fifth YDFA-stage (rode-type) we combined the optical output power of five multi-mode diodes. 

To couple respectively separate the signal and pump light going into respectively coming out of the Ytterbium fibers we used wavelength division multiplexing (WDM) elements. For the first three fiber amplification stages we used all-in fiber filter WDMs. Note that we didn't use mode-field adapters for these three stages. The LMA and rod-type YDFA-stages contained free-space laser beam propagating parts and we used dichroic mirrors as WDM elements.

Efficient free-space single mode signal coupling in between the amplification stages was achieved by carefully choosing pairs of the collimating and coupling lenses in order to match the numerical apertures of the fibers. The collimating lens for the pump light was chosen in a way that the pump light was coupled with high numerical apertures into the pump cladding of the respective fiber.

In order to further reduce the amount of ASE we introduced two bandpass spectral filters in between the first three amplification stages. In between the first and second amplification YDFA-stage we used a 3.7~nm full width half maximum (FWHM) bandpass filter and in between the second and third YDFA-stage a 2.8~nm FWHM bandpass filter.


\section{Results} \label{sec:sections}
The fiber coupled spectral output characteristics of our gain-switched laser diode is shown in Figure~\ref{fig:seed}. The maximum output power is at 1063~nm and the full spectral bandwidth is 0.7~nm. It provides a fiber coupled seed power of 13~$\upmu$W at a fixed repetition rate of 40.03$\pm0.15$~MHz that is set by the driving electronics. The pulse energy was 0.33~pJ. We could not measure directly the pulse duration with our autocorrelator, therefore we measured the pulse duration after the first YDFA-stage at low gain values by replacing the dispersion compensating element by a zero dispersion element. The measured autocorrelator-trace FWHM was FWHM$_{\mathrm{ac}}=$50.9~ps corresponding to an effective pulse width of FWHM$_{\mathrm{Sech^2}}=$33~ps, assuming a hyperbolic secant pulse shape. 

We characterized the polarization state of the free space output of the gain-switched laser diode by recording the angle-dependent power transmission through polarizing beam splitter cubes (PBS) as shown in Figure~\ref{fig:seed}b. By inserting an additional quarterwave-plate before the light passed the polarizing beam splitter cube we could distinguish between circular and unpolarized state and we found the output to be unpolarized.

Each YDFA-stage was optimized towards optimal fiber length for efficient pump absorption, high gain and low ASE contribution. 

The first YDFA-stage was gain-limited due to occurrence of spontaneous lasing at a wavelength around the Ytterbium gain maximum at 1030~nm. We assume that the spontaneous lasing happened because of back-reflections around 1030~nm in the dispersion compensating element. 

\begin{figure}
   \begin{center}
   \begin{tabular}{c c}
   \includegraphics[width=7.4cm]{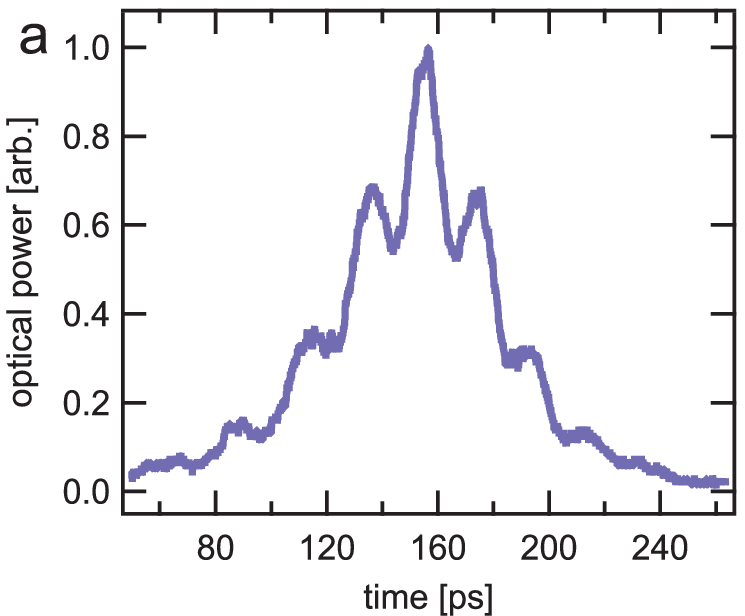} & \includegraphics[width=7.4cm]{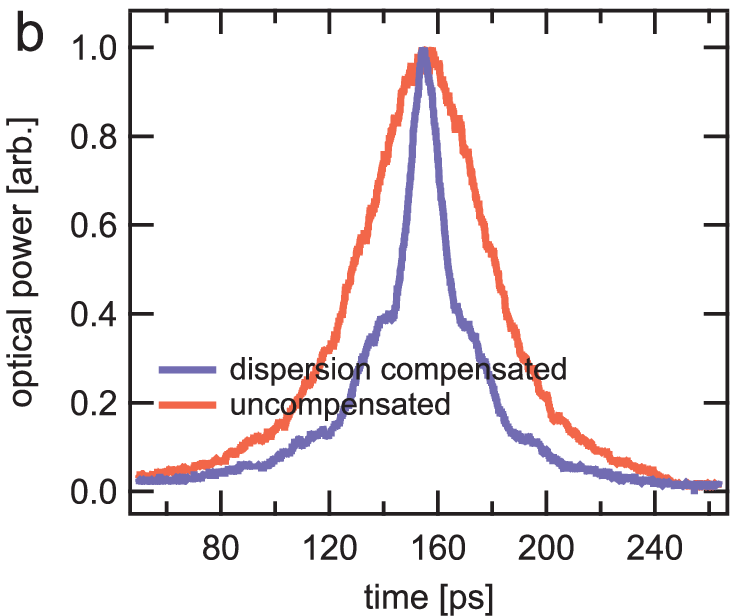} 
   \end{tabular}
   \end{center}
   \caption[example] 
   { \label{fig:pulse_shapes} The two graphs show the two typical autocorrelator traces that were observed by adjusting the polarization controllers. The blue traces show the traces with a dispersion compensating element included in the first YDFA-stage. The red trace shows the recorded trace recorded with no dispersion compensation. }
   \end{figure}

An important step is the linear polarization of the unpolarized beam after the first YDFA-stage. We found that depending on the chosen polarization plane, different pulse shapes were observed for the polarized beam after the in-line polarizer. In Figure~\ref{fig:pulse_shapes} (blue lines) are shown the two typical autocorrelator traces of the linear polarized pulses. For the polarization plane chosen in Figure~\ref{fig:pulse_shapes}a we observed multiple pulses, whereas in the polarization plane chosen in Figure~\ref{fig:pulse_shapes}b we observed single pulses. For comparison purposes we replaced the dispersion compensation optical element by a zero dispersion element and recorded the autocorrelator traces plotted as red line in Figure~\ref{fig:pulse_shapes}b. This visualizes the compression of the pulse duration by a factor of approximately 3 by the dispersive element.

   \begin{figure}
   \begin{center}
   \begin{tabular}{c c}
   \includegraphics[width=7.4cm]{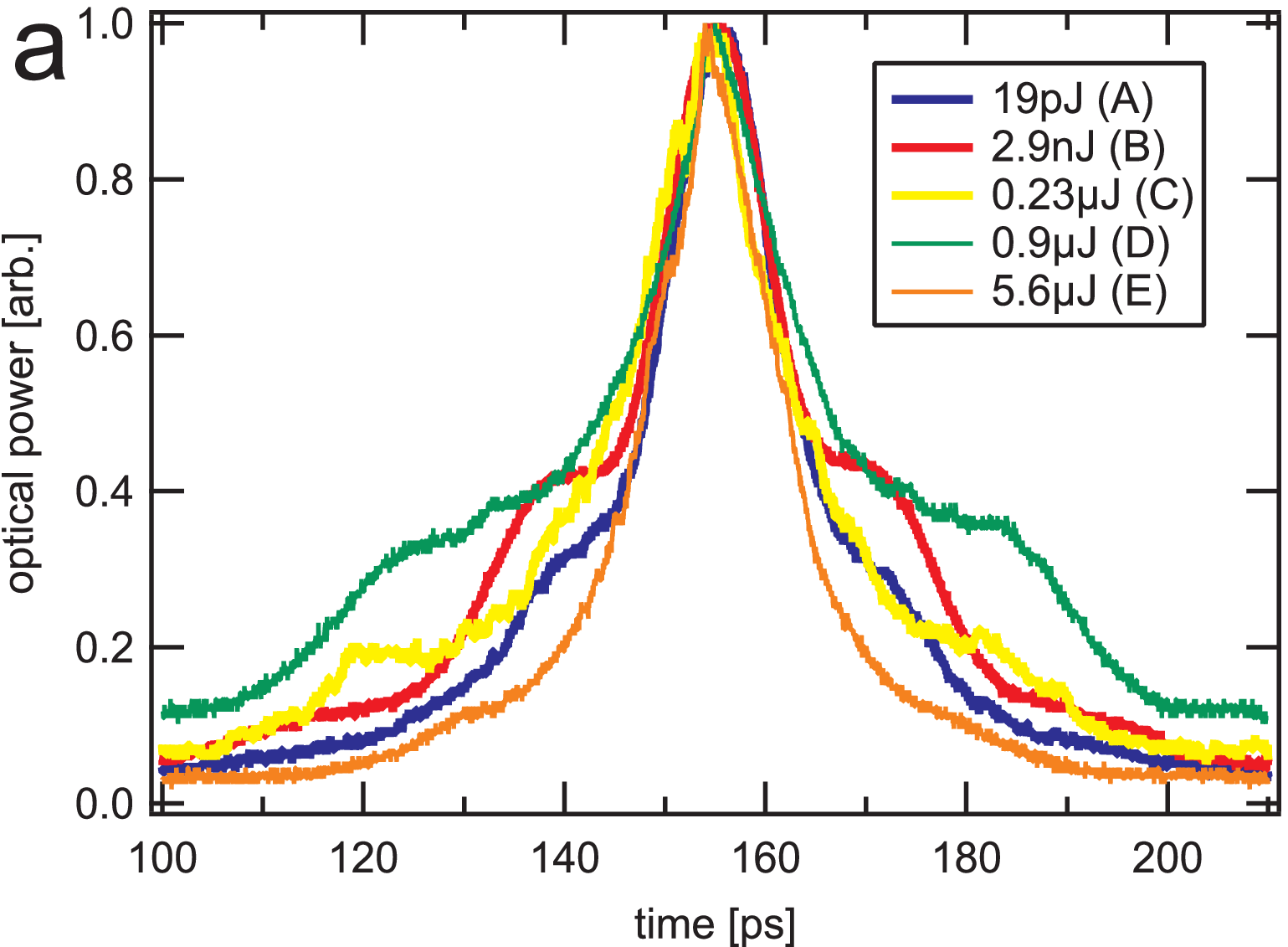} & \includegraphics[width=8.5cm]{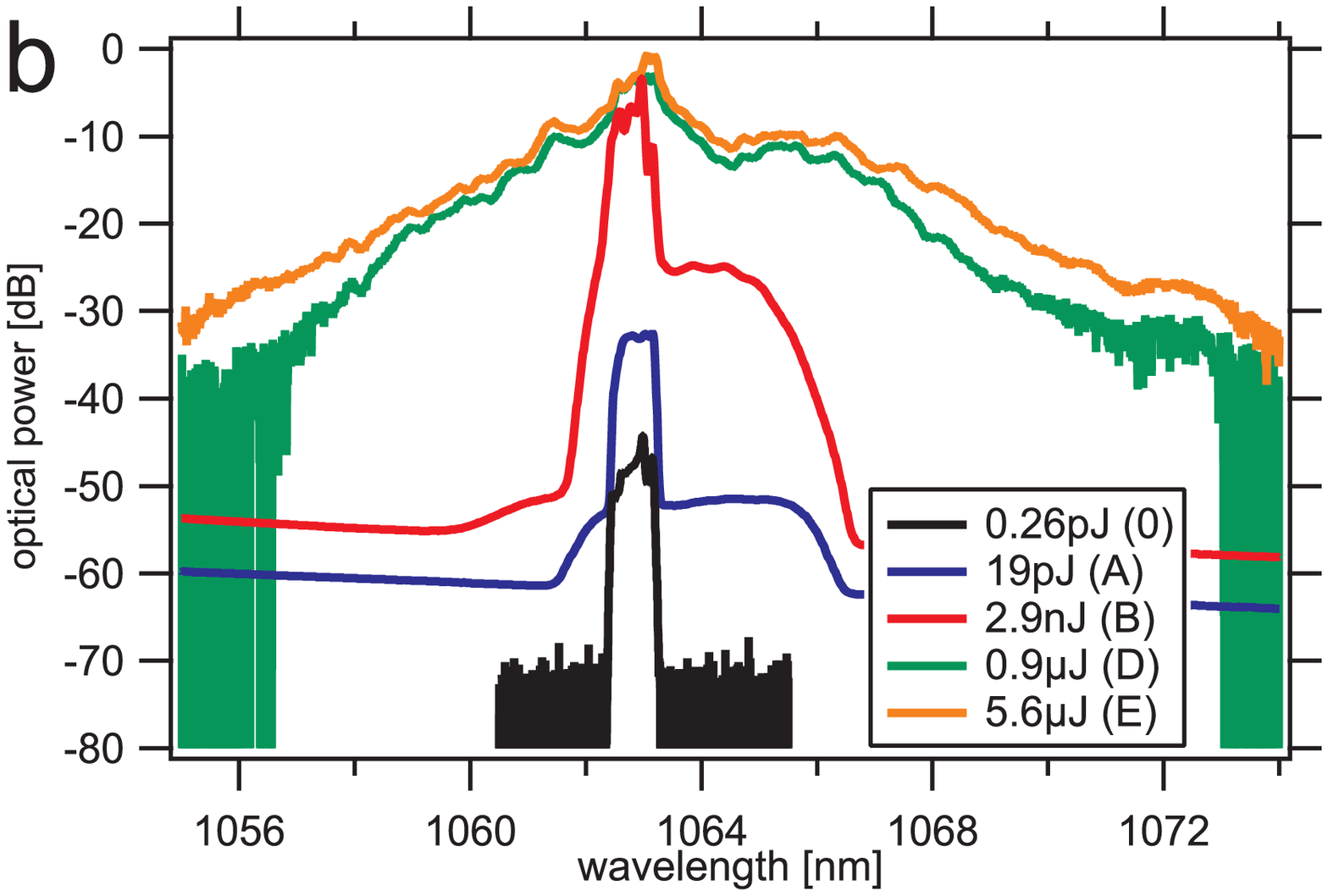} 
   \end{tabular}
   \end{center}
   \caption[example] 
   { \label{fig:pulseshapes} a) Recorded autocorrelator traces at the output of each YDFA-stage. The position of measurement is indicated in the sketch of the experimental setup (Figure~\ref{fig:experimental_setup}) with the capital letters in brackets given in the legend of this graph. Note that the variation of the side lobe is due to slightly varying settings of the polarization controllers and thus the polarization plane passing the in-line polarizer. b) Spectras recorded for each YDFA-stage. Spectral broadening was observed. The black curve shows the spectrum of the gain-switched laser diode at (0) in Figure~\ref{fig:experimental_setup}.}
   \end{figure}

The second YDFA-stage was gain-limited by the occurrence of temporal pulse width broadening. After the second YDFA-stage the repetition rate was reduced from 40~MHz to 1~MHz by using a pulse-picker. The achievable gain while keeping short pulse duration of the third YDFA-stage was also gain-limited by the occurrence of pulse broadening. For both stages we observed a rising peak at approximately 1217~nm. This is in the wavelength region, where a Raman-peak is expected to arise for such fibers. 

The fourth (LMA) YDFA-stage was gain-limited by the occurrence of spontaneous lasing. 

In the fifth (rod-type) YDFA-stage we achieved an average signal output power 5.61W at 1~MHz repetition rate. This corresponds to a pulse energy of 5.7~$\upmu$J and a peak power of $>$0.6~MW. The gain was limited to available pump power.

\begin{figure}
   \begin{center}
   \begin{tabular}{c c}
   \includegraphics[width=8.6cm]{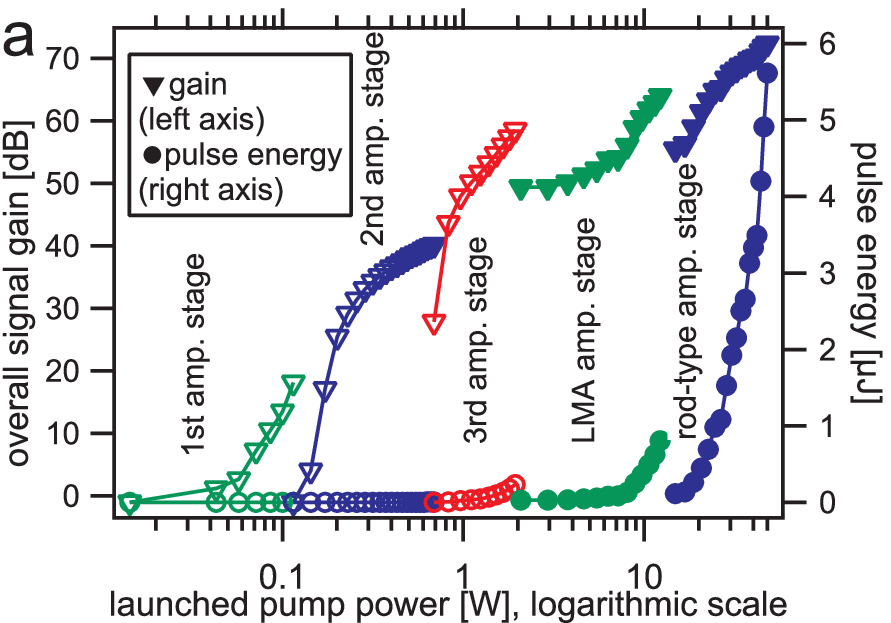} & \includegraphics[width=7.5cm]{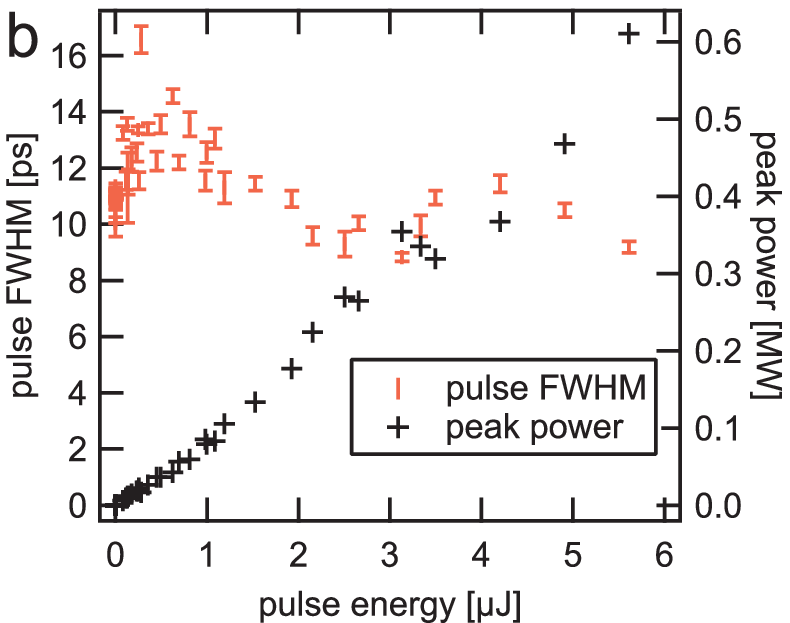}  \\
   \end{tabular}
   \end{center}
   \caption[example] 
   { \label{fig:pulseE_vs_pulseFWHM_peakpower} a) Amplification characteristics of all five stages. The graph shows the output measured after each YDFA-stage. As seed for the next YDFA-stage was used the  output of the preceding YDFA-stage. Note that the gain-curves respectively the pulse energy curves of a later YDFA-stage starts below the preceding output because in between the stages were placed optical elements (c.p. Figure~\ref{fig:experimental_setup}) reducing the transmitted power.  b) Measured pulse energies and pulse durations (we assumed sech$^2$-pulse shape) of the amplified pulses over all stages. From these values the peak power was estimated.}
   \end{figure}

In Figure~\ref{fig:pulseE_vs_pulseFWHM_peakpower} is shown the amplification characteristics over all five stages. The achieved total gain was 72~dB resulting in a pulse energy of 5.7~$\upmu$J at a pulse duration of $<$11~ps FWHM. The final peak power was estimated to be $>$0.6~MW. The pulse shapes did not change significantly throughout the amplification process as indicated by the autocorrelator traces shown in the Figure~\ref{fig:pulseshapes}b. These traces were recorded at the output of each YDFA-stage. We observed spectral broadening by a factor of approximately 5 from 0.7~nm (full width) to 3.4~nm (full width), as shown in Figure~\ref{fig:pulseshapes}b. From the spectras shown in Figure~\ref{fig:pulseshapes}b we estimated the  the signal to ASE ratio to be approximately 20~dB for all amplification stages.

\section{Conclusion} 
The output of a gain-switched laser diode was compressed by a factor of approximately 3 by using a dispersive optical element to a pulse duration of 11~ps.

We have demonstrated the amplification of the weak pulses (0.33~pJ) by 72~dBs to an output pulse energy of 5.7~$\upmu$J. No significant temporal pulse broadening was observed during the amplification process. We estimated the peak power to be $>$0.6~MW. 

To the best of our knowledge, we achieved the shortest pulse duration and highest peak power with a MHz repetition rate driven gain-switched diode based fiber amplifier system.

\appendix    

\acknowledgments     
This study was financed in parts by the Swiss Commission for Technology and Innovation (CTI 11196.1).
 


\bibliography{library}   
\bibliographystyle{spiebib}   

\end{document}